# Object-Oriented Approach for Integration of Heterogeneous Databases in a Multidatabase System and Local Schemas Modifications Propagation

Mohammad Ghulam Ali

Indian Institute of Technology, Kharagpur
Academic Post Graduate Studies and Research
Kharagpur, India
ali_iit@yahoo.com, ali@hijli.iitkgp.ernet.in

*Abstract -* **One of the challenging problems in the multidatabase systems is to find the most viable solution to the problem of interoperability of distributed heterogeneous autonomous local component databases. This has resulted in the creation of a global schema over set of these local component database schemas to provide a uniform representation of local schemas. The aim of this paper is to use object-oriented approach to integrate schemas of distributed heterogeneous autonomous local component database schemas into a global schema. The resulting global schema provides a uniform interface and high level of location transparency for retrieval of data from the local component databases. A set of integration operators are defined to integrate local schemas based on the semantic relevance of their classes and to provide a model independent representation of virtual classes of the global schema. The schematic representation and heterogeneity is also taken into account in the integration process. Justifications about Object-Oriented Modal are also discussed. Bottom up local schema modifications propagation in Global schema is also considered to maintain Global schema as local schemas are autonomous and evolve over time. An example illustrates the applicability of the integration operator defined.**

## I. INTRODUCTION

To efficiently retrieve information from heterogeneous and distributed data sources has become one of the top priorities for institutions across the business world. Information from these sources needs to be integrated into one single system such that the user can retrieve the desired information through the integrated system by a single query.

Heterogeneous data source means there is no homogeneity among the databases at various sites or at least component databases differ in some important respect (e.g. the DBMS they are running or perhaps the data model implemented by it – relational, object-oriented, etc)

A multidatabase system (MDBS) is a database system that resides on top of existing local autonomous component databases systems (CDBSs) and presents a single database to its users [1]. MDBS usually maintains a single global database schema [1], which is integration of all component database schemas. The schema contains the format, structure, organization of the data in a system. MDBS maintains only the global schema and the CDBS actually maintains all user data. Creating and maintaining the global schema, which requires the use of database integration techniques, is a critical issue in multidatabase systems. A variety of approaches to schema integration have been proposed; e.g., [4,5,6,7]. The main schema integration problem is associated with combining diverse schemas of the different databases into a coherent global view by reconciling any structure or semantics conflicts between the local component databases [14].

In this paper, we have defined an integration methodology with resolving conflicts and maintaining transparency. The conflicts are fundamental as different schemas will be developed under different data models and ideology. This results in both structural and semantic conflicts between schemas which must be resolved to construct global schema.

Transparency is important as the integration should be performed transparently to the user and the application programmer.

## Generic framework for integrating heterogeneous component database systems

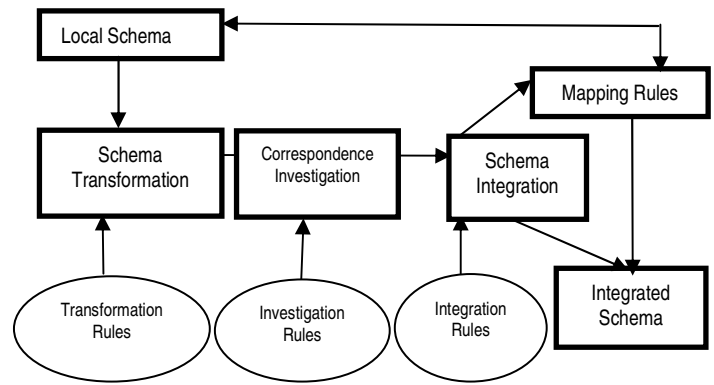

Figure 1

Using this techniques (figure 1), the multidatabase system's designer usually transform the local schemes into a common data model, compare local schemas to exploit their semantic correspondences and identify their schematic differences, merge the local schemas into global schema (or a set of global schemas), and define the mappings between the global schema and the local schemas.

### Why we have chosen Object-Oriented Data Model?

In our approach, we use the object-oriented data model as common data model to construct a global conceptual model by set of integration operators. The object-oriented modal has grown in popularity because of its simplicity in representing complicated concepts in an organized manner or we say due to its semantic richness and availability. We can also say that in object-oriented data model there is only one concept (objects) is used as opposed to the two concepts (entities and relationship) of the entity relationship modal. Using two concepts may cause problems in the process of schema integration and schema transformation because the same real world object may be modeled as entity in one schema but as a relationship in another schema [17]. Using encapsulation, methods, and inheritance, it is possible to define almost any type of data representation. Consequently, object-oriented models have been used not only to model the data, but also to model the data models. That is, object-oriented models have seen increased use as canonical models into which all other data models are transformed into and compared. The object-oriented model has very high expressiveness and is able to represent all common modeling techniques present in other data model. Hence, it a nature choice for a canonical model [14]. In general, tightly coupled systems integrate the diverse sources of intelligence through a global conceptual scheme, normally denominated canonical model, providing a uniform vision of the diverse component at a high level [15]. The use of a canonical model hides the structural differences between the different components and gives to the user the illusion to be accessing a simple centralized database. Since we adopt an object-oriented data model as the data model of the multidatabase system [4,5,6,7,8] also illustrated in figure 2. The global schema as we propose in an object-oriented data model is a virtual schema because no actual data are stored for this schema. Thus, the classes in the global schema are called virtual classes and the objects associated with the classes are called virtual objects.

Where

GDM – is local schema in Global Data Model
LCS – is logical organization at each site
LIS – is physical data organization on each site

**Schemas transformation** is a process to transform heterogeneous database schemas into the global common data model and also to support query expressed in the global common data manipulation language have to be decomposed & translated into the local query language and also query produces a response, the sub-results coming from individual local schemas are then to be transformed and composed into the format of multidatabase system response language (as shown in figure 2), See for details the work of [4,6,8,16,17]. A schema transformation should only be applied to restructure a schema if the resulting schema is equivalent to the original schema.

**Correspondence assertions** for the database administration (DBA) to specify the semantic correspondence among component schemas as explained in [2,5,6]. Correspondence assertions mean correspondence between tables and classes, correspondence between columns and attributes. Based on these assertions, integration rules are designed. See the work of [5,6,7,9], which use a set of primitive integration operators to do the integration. See the work of [5]. Various restructuring operators are proposed to rename or restructure schema objects of component schemas to resolve conflicts between them.

The **mapping information** is important for the decomposition of a global query against global schema and for this mapping strategy between the global schema and component schemas is necessary i.e. mapping for attributes and object instances among a virtual class and its constituent classes (classes in the component schemas). See for details, the work of [4,7].

As local component databases are autonomous likely to evolve overtime by restructuring their schemas, the schema evolution and schema modifications propagation are also an important issues in the multidatabase system [1,6,10,11,12]. We have taken into consideration this issue in this paper. Since schema evolution is an incremental process where schema of a database is subject to change due to a variety of reasons e.g. structural changes in the organization and courses to be reflected in the database, design mistakes in the old database need to be corrected, few researchers address the problem of schema evolution in a multidatabase system [10,12]. The methodology to propagate the effects of restructuring local schemas to the integrated global schema without having to perform the integration process from the beginning, few researchers have also addressed this issue [6,11].

In this paper, we propose an approach to database integration between heterogeneous databases and during this process; we take into consideration of the *system heterogeneity*, *physical heterogene*ity (syntactic heterogeneity) and the *Schematic* & *semantic heterogenei*ty [13].

The **system heterogeneity** concerns the differences in the low level architectural platforms local component databases, such as hardware configuration, operating systems and communication facilities. As sites might have different systems techniques used by different DBMSs such as query processing strategy, concurrency control mechanism and transaction management facilities.

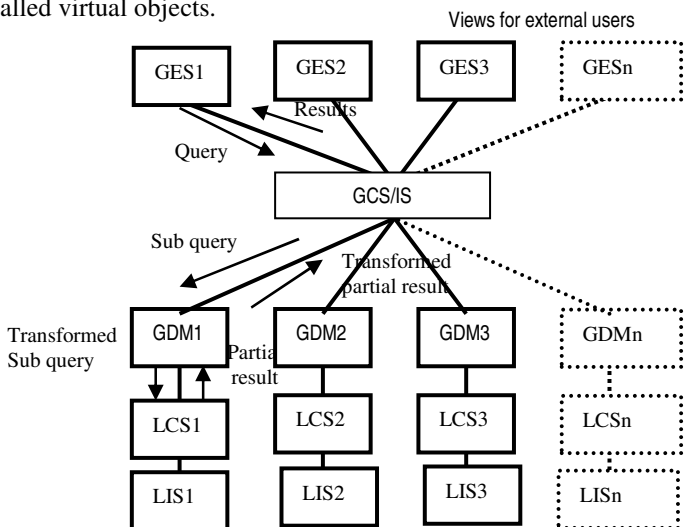

Figure 2

The **physical heterogeneity/syntactic heterogeneity** refers to the differences in the data representation. It may concern the encoding of same data values into integers or real of different types. Next, it may concern different implementation of the data model, e.g., of tables in different relational systems. The formats of manipulation procedures, e.g., of SQL queries, also often differ. Finally, the data models themselves can differ. Data of interest may be in a relational database, and in an object-oriented database.

The **Schematic & semantic heterogeneity** refers to the discrepancies between data names, values, and the conceptual structures.

We first define schematic & semantic heterogeneity**,** semantic relationships between the component databases including attributes correspondence – these determine how local classes should be integrated in the global schema. We then present integration operators that provide steps for constructing an integrated schema from two component databases according to semantic relevance and attributes correspondence. These operators achieve integration via merging, generalizing, specializing, or importing local classes to the global schema.

Section 2 explains correspondence Investigation (semantic & schematic heterogeneity, semantic relevance/correspondence between classes). Section 3 explains construction of integrated schema. Section 4 Describes local schemas modifications propagation into a global schema and maintenance of Global Schema. Section 5 Illustrates work by examples. Section 6 concludes the work.

## II. CORRESPONDENCE INVESTIGATION

### A. Schematic & Semantic Heterogeneity

**Type of heterogeneity**

**Data heterogeneity:** refers to difference among local definitions, such as attribute types, formats, or precision [2].

**Schematic & Semantic Heterogeneity: Schematic** heterogeneity means, all component databases have different data model structure where equivalent and related data concepts are present and having conflicting structural representation. We can also say all local component databases model having same real world concept in a structurally different manner. **Semantic** refer to the meaning of data, which only defines the structure of the schema items (e.g., classes and attributes). Semantic heterogeneity refers to differences or similarities in the meaning of local data. For example, two schema elements in two local data sources can have the same intended meaning, but different names. Thus, during integration, it should be realized that these two elements actually refer to the same concept. Alternatively, two schema elements in two data sources might be named identically, while their intended meanings are incompatible. Hence, these elements should be treated as different things during integration.

It is vital to compare local schema contents to detect type of schematic conflicts that may exist among these local schemas [2,3,4,6,7]. Schematic heterogeneity arises when information that is represented as data under one schema, is represented within the schema in another, for example as relation or class names.

### B. Semantic relevance / correspondence between classes

Based on their real world semantics, schematic representations and extent classes can be related to each other in one of the following 4 semantic relationships – these determine how local classes should be integrated in the global schema [5,6].

Let S1 and S2 be local schemas, C1 and C2 be local classes, type() is a function that returns the set of class, and ext() is a function that returns the extent (i.e. the set of objects) of a class.

**Equivalence**

A class C1∈ S1 is said to be equivalence to class C2∈ S2 if

- Represents same real world concept

- Same schema name

- Difference in number of attributes and data types

The extents of S1-employee and S2-Employee could be identical (S1=S2), disjoint (S1 ∩ S2 = φ) or overlapping (S1 ∩ S2 ) ≠ φ)

**Synonymy**

Classes have the same definition as equivalent classes except that the names of class C1 and C2 are synonymous (different name)

**Containment**

A class C1∈ S1 is said to be contained in a class C2∈ S2 if

- Type(C1) ⊃ Type(C2)
- Ext(C1) ⊂ ext(C2)
- C1 is a subclass of C2

**Homonymy**

Two classes are said to be homonymous if they represent different real world concepts and have the same name. These classes cannot be merged into a common virtual class in the global schema.

Semantic relationship is defined at the attribute level: correspond-to (≡). Several attributes are said to correspond to each other and therefore can be represented as one global attribute.

Example

S1-employee(empno, name, address, phone, salary)
S2-employee(number, name, address, dob, phone, salary)
empno ≡ number,
name ≡ name,
address ≡ address,
phone ≡ phone,
salary ≡ f(salary)

Where f is a conversion function that convert salaries in one currency type to common global type.

## III. CONSTRUCTION OF INTEGRATED SCHEMA / GLOBAL SCHEMA

Global Schema/Integrated Schema is a collection of virtual classes constructed from a set of local classes. In our integration methodology, integration operators are applied to set of local classes to generate virtual classes. The choice of integration operator depends on semantic relationship that exists between local classes.

Pre-process of the schema integration is identifying semantic relationship & attributes correspondence and then restructuring schema objects of component schemas to resolving conflicts between them. We have already discussed on these issues.

We propose a set of class integration operators to generate a virtual class as global class for an integrated schema in multidatabase.

Each Virtual Class in Global Schema consists of four elements

   i.        Integration Operators,
   ii.       Set of Local Classes,
   iii.      Set of Attributes in Virtual Classes and
   iv.      Some Integration Rules such as mapping virtual class in global schema to various local classes in component schemas

### A. Union/Combine/Merge Operator

The Union operator integrates class1 from DB1 and class2 from DB2 into a virtual class as new class. Only new class will appear in the Global Schema. It merges equivalent or synonymous local classes into a single virtual class. It has the following syntax:

Union(class1 from DB1, class2 from DB2, virtual class)

For n numbers of local databases

Union(class 1 from DB1, Class 2 from DB2, Class n from DBn, virtual Class)

Where virtual attributes of the virtual class are
$$VC_a = C1_a \cup C2_a \cup \ldots \cup Cn_a$$
and virtual objects of virtual class are
$$VC_o = C1_o \cup C2_o \ldots \cup Cn_o$$

### B. Generalize Operator

The generalize operator creates a common superclass (as shown in figure 3) of class1 from DB1 and class2 from DB2. Local classes have to be equivalent, synonymous or contained-in classes. It has the following syntax:

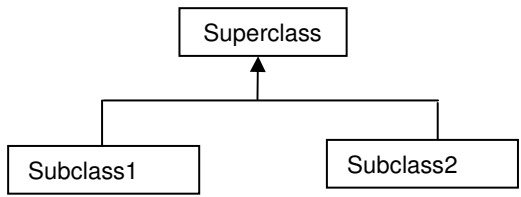

Figure 3. Generalization Inheritance

Generalize(class1 from DB1, class2 from DB2, virtual class)

For n numbers of local databases

Generalize(class 1 from DB1, Class 2 from DB2, Class n from DBn, virtual Class)

Where virtual attributes of the virtual class are
$$VC_a = C1_a \cap C2_a \cap \ldots \cap Cn_a$$
and virtual objects of virtual class are
$$VC_o = C1_o \cup C2_o \cup \ldots \cup Cn_o$$

We show an example through figure 4 for generalizing two subclasses into a superclass.

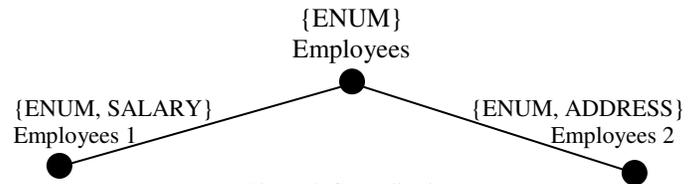

Figure 4. Generalization

It is the reverse of the specialization process.

### C. Specialize Operator

The specialize operators creates a common subclass (as shown in figure 5) of class1 from DB1 and class2 from DB2. Local classes have to be equivalent, synonymous or contained-in classes. It has the following syntax:

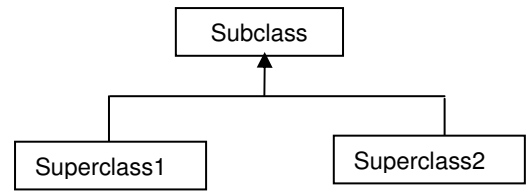

Figure 5. Specialization Inheritance

Specialize(class1 from DB1, class2 from DB2, virtual class)

For n numbers of local databases

Specialize(class 1 from DB1, Class 2 from DB2, Class n from DBn, virtual Class)

Where virtual attributes of the virtual class are
$$VC_a = C1_a \cup C2_a \cup \ldots Cn_a$$
and virtual objects of virtual class are
$$VC_o = C1_o \cap C2_o \cap \ldots \cap Cn_o$$

We show an examples through figure 6 for specializing two superclasses into a subclass or may be more subclasses.

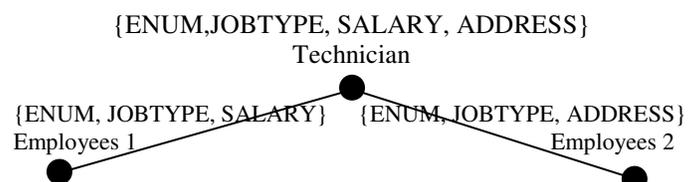

Figure 6. Specialization

We explain more about specialization as it is is a process of defining a set of subclasses of a superclass. For example {SECRETARY, ENGINEER, TECHNICIAN} is a specialization of EMPLOYEE superclass based upon job type.

May have several specializations of the same superclass or more than one superclasses.

### D. Import Operator

Import operator import a class from a local schema without merging them with class from other local schemas. It is a straight forward mapping

Import operator has syntax:

Import(Class from DB1)
**VC1 = C1**
Where virtual attributes of the virtual class are
**$VC1_a = C1_a$**
and virtual objects of the virtual class are
**$VC1_o = C1_o$**

## IV. LOCAL SCHEMAS MODIFICATIONS PROPAGATION INTO A GLOBAL SCHEMA AND MAINTENANCE OF GLOBAL SCHEMA

Many researchers have addressed issues of local schemas transformation, schema mapping, resolving of schema schematic & semantic heterogeneity, generating common global data modal, identifying class and attributes correspondence and creating global schema based on common global data modals. Few researchers have addressed the issue of local schemas modification propagation into a global schema.

We propose a method through which we can maintain global schema without manual intervention.

We show a proposed diagram through Figure 7.

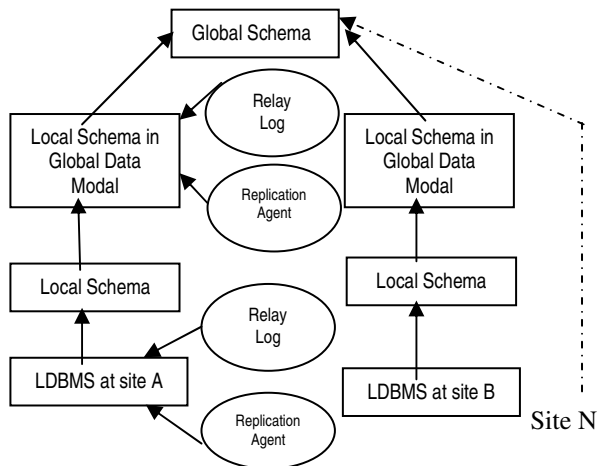

Figure 7. Local Component Database Modification Propagation

Now the main challenging issue is now how to maintain global schema as all local schemas are autonomous and evolve over time. How to automatically propagate local schemas modifications into a global schema. How to update Global Directory or Global Dictionary or we say metadata information of Global Schema based on local schemas. The maintenance of this metadata can become an issue, as the local schemas evolves over time, there should be a mechanism to import and synchronize metadata for global schema

If every local heterogeneous database management system has data model independent replication agent that will replicate/relay schema modifications in bottom up unidirectional manner to Global Data Modal and similarly if all Global Data Modals have the same replication agent that will replicate/relay in the same manner every schema modifications to global schema then the global schema will be maintained automatically without manual intervention and the Global Directory will be maintained. In case, if any local site global data modal is not working at the time of local schema modifications then all local schema modifications will be stored in a schema update query log and all pending schema modifications queries log will be relayed as soon as the same site gets up. This method will only take care of relay schema modifications not data as the replication user that is used by the replication agent will have only alter privilege permission to alter schema in bottom-up manner.

## V. ILLUSTRATE OUR WORK BY AN EXAMPLE

*Example 1*

Class employees in DB1 at Site A

| employeecode | name | country | age |
|---|---|---|---|
| 1 | john | NY | 25 |
| 2 | peter | NY | 26 |
| 3 | albert | NY | 27 |

Class employees in DB2 at Site B

| employeecode | name | country | age | phone | age |
|---|---|---|---|---|---|
| 4 | habib | IN | 25 | 28789 | 25 |
| 5 | mohan | IN | 28 | 22789 | 28 |
| 6 | mary | IN | 29 | 23789 | 29 |

In the above example, both classes are of equivalence nature. Now we combine both classes using our proposed union integration operator to form a virtual class as employees in the integrated/global schema where attributes of the virtual class are union of attributes of both classes and objects of virtual class are union of objects of both classes. When we will submit query on resulting virtual class will get result as follows:

Employees

| employeecode | name | country | age | phone |
|---|---|---|---|---|
| 1 | john | NY | 25 | |
| 2 | peter | NY | 26 | |
| 3 | albert | NY | 27 | |
| 4 | habib | IN | 25 | 28789 |
| 5 | mohan | IN | 28 | 22789 |
| 6 | mary | IN | 29 | 23789 |

.

Class UGStudents in DB1 at Site A

| Id | name | country | CGPA |
|---|---|---|---|
| 1 | john | NY | 9 |
| 2 | peter | NY | 8 |
| 3 | albert | NY | 7 |

Class PGStudents in DB2 at Site B

| Id | name | country | CGPA | Age |
|----|------|---------|------|-----|
| 4 | habib | IN | 9.6 | 20 |
| 5 | mohan | IN | 8.6 | 22 |
| 6 | mary | IN | 7.6 | 23 |

In the above example, both classes are of synonymous nature.

Now we combine both classes using our proposed generalize integration operator to form a virtual class, as persons in the integrated/global schema where attributes of the virtual class are intersection of attributes of both classes and objects of virtual class are union of objects of both classes. When we will submit query on resulting virtual class will get result as follows:

Persons

| id | name | country | CGPA |
|----|------|---------|------|
| 1 | john | NY | 9 |
| 2 | peter | NY | 8 |
| 3 | albert | NY | 7 |
| 4 | habib | IN | 9.6 |
| 5 | mohan | IN | 8.6 |
| 6 | mary | IN | 7.6 |

GStudents and PGStudents classes have containment relation with person.

*Example 3*

Class teachers in DB1 at Site A

| Id | name | country | designation |
|----|------|---------|-------------|
| 1 | john | NY | Asst Prof |
| 2 | peter | NY | Assoc Prof |
| 3 | albert | NY | Prof |

Class teachers in DB2 at Site B

| Id | name | country | designation | DOB | DOJ |
|----|------|---------|-------------|-----|-----|
| 4 | habib | IN | Asst Prof | 27/01/68 | 10/08/1999 |
| 5 | mohan | IN | Assos Prof | 28/01/68 | 11/08/1999 |
| 6 | mary | IN | Prof | 29/01/68 | 12/08/1999 |
| 3 | albert | NY | Prof | 30/01/68 | 13/08/1999 |

In the above example, both classes are of equivalence nature. Now we combine both classes using our proposed specialize integration operator to form a virtual class as professor in the integrated/global schema where attributes of the virtual class are union of attributes of both classes and objects of virtual class are intersection of objects of both classes. When we will submit query on resulting virtual class will get result as follows:

professor

| Id | name | country | designation | DOB | DOJ |
|----|------|---------|-------------|-----|-----|
| 3 | albert | NY | prof | 30/01/68 | 13/08/1999 |

## VI. CONCLUSION

This paper proposes an approach to database integration and creation of global database as group of homogeneous virtual classes from heterogeneous local classes. In our approach, correspondence investigations between schema objects of component databases are specified first. Then integration operators are applied to construct the integrated schema. We have also discussed how to propagate local schema modifications propagation into a global schema without manual intervention. We have illustrated work by examples. In the future, we plan to address other issues involved during schema integration.

## AUTHORS PROFILE

**Mohammad G. Ali** I am a System Engineer Grade I in the Indian Institute of Technology Kharagpur, West Bengal, India. I am associated with System Analysis & Design, Programming, Implementation and Maintenance of Client-Server DBMSs and Web Application Developments. I am also associated with Database Administration, Web Server Administration, System Administration and Networking of the Institute. I have deployed many small to big projects on our Institute Network.